%% file: eprint.tex
\documentclass[10pt]{article}
\usepackage{graphicx}

\def\Title#1{\begin{center} {\Large #1 } \end{center}}
\def\Author#1{\begin{center}{ \sc #1} \end{center}}
\def\Address#1{\begin{center}{ \it #1} \end{center}}

\newcommand\pubblock{\rightline{\begin{tabular}{l} Proceedings of the Fifth Annual LHCP\\ \pubnumber\\
         \pubdate  \end{tabular}}}

\newenvironment{Abstract}{\begin{quotation} \begin{center} 
             \large ABSTRACT \end{center}\bigskip 
      \begin{center}\begin{large}}{\end{large}\end{center} \end{quotation}}

\newenvironment{Presented}{\begin{quotation} \begin{center} 
             PRESENTED AT\end{center}\bigskip 
      \begin{center}\begin{large}}{\end{large}\end{center} \end{quotation}}

\input econfmacros.tex

\usepackage{color}

 \newcommand\pubnumber{ }

\newcommand\pubdate{\today}

\def\affiliation{
On behalf of the ATLAS, CMS, and LHCb Experiments, \\
Department of Physics \\
National Taiwan University, Taipei, Taiwan}

\begin{document}

\large
\begin{titlepage}
\pubblock

\vfill
\Title{  Rare Decays  }
\vfill

\Author{ Kai-Feng Chen  }
\Address{\affiliation}
\vfill
\begin{Abstract}

Studies of rare decays play an important role in the search of physics beyond the standard model.
New particles may participate in the loop processes and can be probed by seeing any 
deviations from the standard model predictions.
The very rare decay $B_s\to\mu^+\mu^-$ has been observed with the data collected by CMS and LHCb experiments. 
The signal seen by the ATLAS experiment is less significant but is compatible with the predictions.  
The measurement itself provides stringent constraints to new physics models. 
The first effective lifetime measurement with $B_s\to\mu^+\mu^-$ candidates has been carried out by the LHCb experiment.
More data are still required to observe the $B^0\to\mu^+\mu^-$ decays. 
The  $B\to K^*\mu^+\mu^-$ decay  also proceeds through a flavour changing neutral current process, and is sensitive to the new physics.
Extended measurements are carried out for $B\to K^*\mu^+\mu^-$ decays. Most of the classical physics parameters are found to be 
consistent with the predictions, but tensions do emerge in some of the observables. 
More data will help to clarify these potential deviations.

\end{Abstract}
\vfill

\begin{Presented}
The Fifth Annual Conference\\
 on Large Hadron Collider Physics \\
Shanghai Jiao Tong University, Shanghai, China\\ 
May 15-20, 2017
\end{Presented}
\vfill
\end{titlepage}
\def\thefootnote{\fnsymbol{footnote}}
\setcounter{footnote}{0}
%

\normalsize 


\section{Rare $B_s \to \mu^+\mu^-$ and $B^0 \to \mu^+\mu^-$ Decays}

The decays $B_s\to\mu^+\mu^-$ and ${B^0}\to\mu^+\mu^-$ are predicted to be tiny in the standard model (SM) since 
these processes can only proceed through flavor-changing neutral currents (FCNC). The FCNC processes are not allowed 
at the leading order. Also, the internal annihilation of quarks in the $B$ mesons is needed and the rate has to be further suppressed. 
Another suppression factor of $m_{\mu}^2/m_{{B}}^2$ ($m_{\mu}$ and $m_{{B}}$ are the masses of the muon and the ${B}$ meson, respectively) is required by the helicity. With the best knowledge to date~\cite{Bobeth:2013uxa}, the branching fractions of these two modes are 
\begin{equation}
\mathcal{B}^{\rm SM}({B}_{s}\to\mu^+\mu^-) = (3.66\pm0.23)\times10^{-9}~{\rm and}~\mathcal{B}^{\rm SM}({B^0}\to\mu^+\mu^-) = (1.06\pm0.09)\times10^{-10}.
\end{equation}

With the help of large production cross section of $b$-hadrons at the LHC, these tiny decay rates become reachable. In particular, several extended theories beyond the SM may enlarge the branching fractions of these two modes. For example, in the MSSM models, the branching fraction 
of  $B_s\to\mu^+\mu^-$ has a strong dependence on $\tan\beta$ to the sixth power. The models which introduce extended Higgs sectors can also increase
the decay branching fractions. In the usual consideration, the ratio of ${B^0}\to\mu^+\mu^-$ and $B_s\to\mu^+\mu^-$ branching fractions
is a good test of the minimal flavour violation scenarios. Any deviation from the SM-based prediction of these two channels is a clear hint of where 
the SM should be extended. However, if the branching fractions are measured to be consistent with the predictions of SM, strong constraints are introduced to several related new physics models, and can even be extended to the scenarios which require an energy above the LHC design. 

The main focus of the analysis is on the background suppression. This is due to the small number of signal events on top of heavy combinatorial background events.
The signal is compatible with a pair of real muons and a displaced decay vertex; the invariant mass of the muon pair should agree with the mass of $B_s$ or $B^0$ mesons. The combinatorial background events are generally consistent with a mixture of muons from the $b$-hadrons semileptonic decays.
These background events can have additional activities other than the candidate muon pair, also, the vertexing quality can be worse as well. Hence these information are included in a boosted decision tree (BDT) analysis to improve the background reduction power.
Another critical background is associated with mis-identified hadrons (mostly kaons or pions) from $B$ meson decays, in particular for  
events with a $B$ meson decaying into two hadrons ($B \to K^+K^-$, $K^+\pi^-$, or $\pi^+\pi^-$). If both hadrons are mis-reconstructed as muons, it forms 
a peaking structure right below the targeted signal ${B^0}(B_s)\to\mu^+\mu^-$. Thus a stringent requirement on the muon identification is needed.

The decay branching fractions of ${B^0}(B_s)\to\mu^+\mu^-$ are normalized by the $B^+ \to J/\psi (\to \mu^+\mu^-) K^+$ decay,
\begin{equation}
\mathcal{B}(B_{s,d} \to \mu^+\mu^-) = {N_{\rm sig} \over N(B^+ \to J/\psi K^+)} \cdot 
\mathcal{B}(B^+ \to J/\psi K^+) \cdot {\epsilon(B^+) \over \epsilon(B_{s,d})}
\cdot {f_u \over f_s},
\end{equation}
where $N_{\rm sig}$ is the signal yield, $\epsilon$ is the detecting acceptance and efficiency, and ${f_u/f_s}$ is the ratio of hadronization fractions.
Similar selection criteria are adopted for the two muons from $J/\psi$ decays, and hence the associated systematic uncertainties can be 
mostly cancelled. LHCb also uses $B \to K\pi$ decays to normalize the events.

\begin{figure}[htb]
\centering
\includegraphics[height=1.5in]{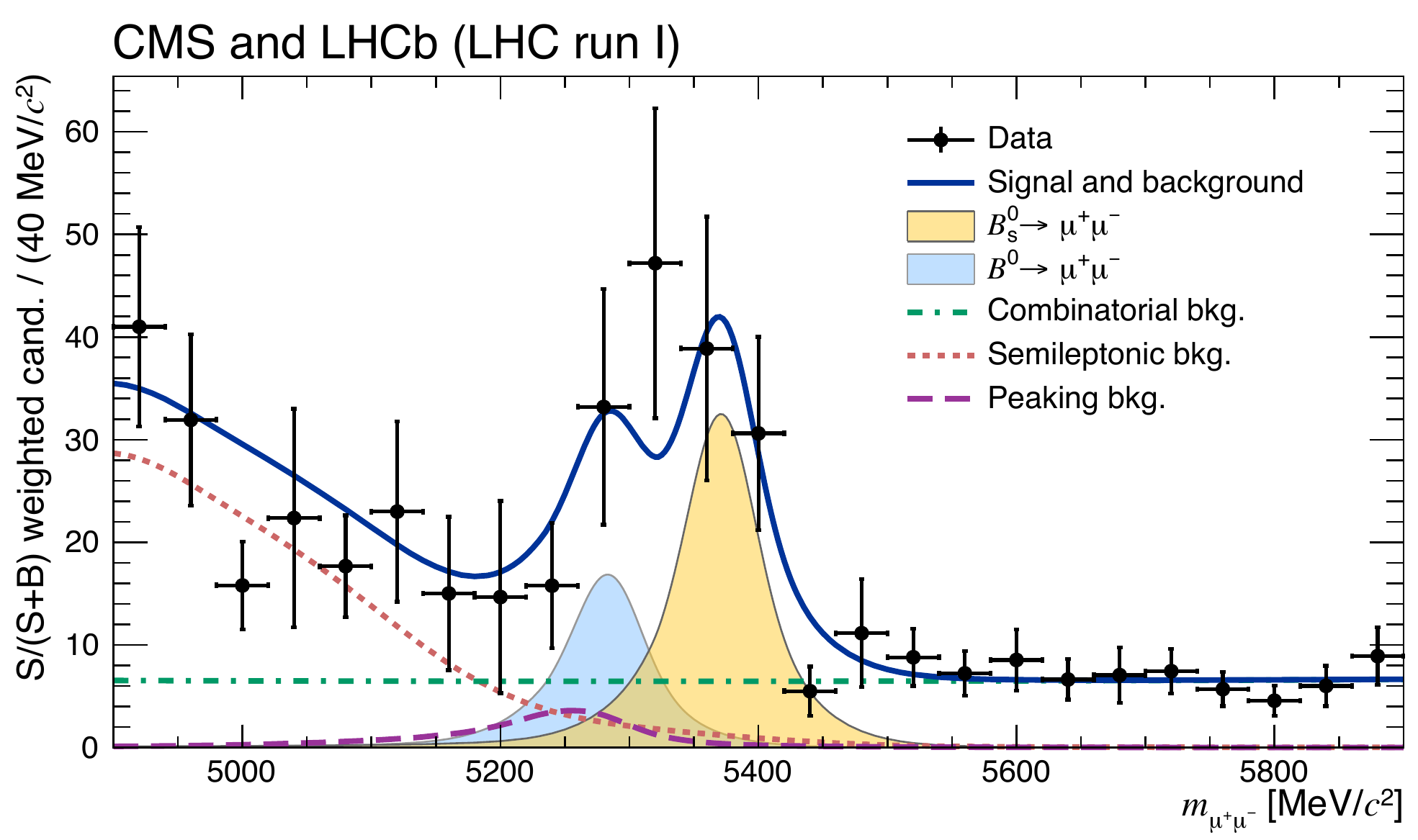}
\includegraphics[height=1.5in]{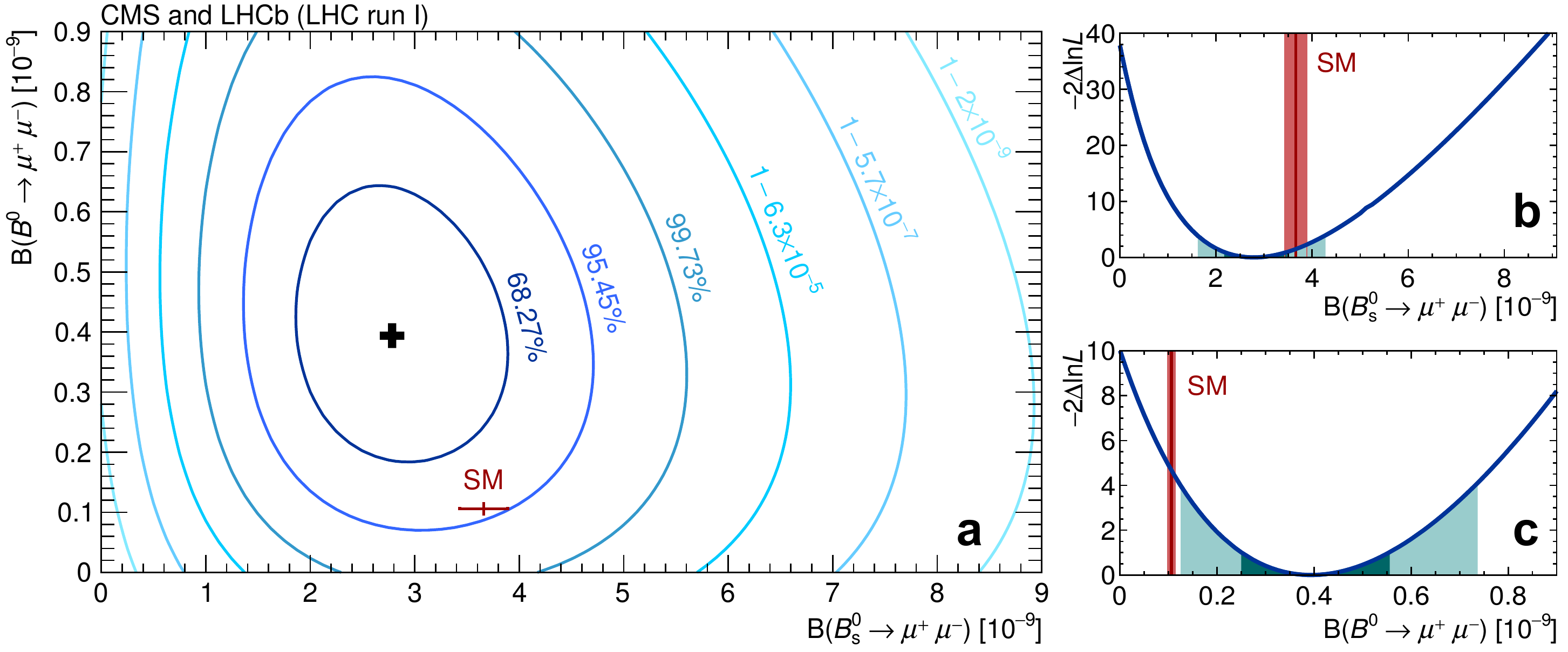}
\caption{ The weighted invariant mass distribution for $B \to \mu\mu$ candidates (left) and the 
results of likelihood scans (right) from the CMS and LHCb combination paper. 
In the likelihood scans, all of the nuisance parameters (systematics) are profiled.}
\label{fig:bmm_combine}
\end{figure}

\begin{figure}[htb]
\centering
\includegraphics[height=2in]{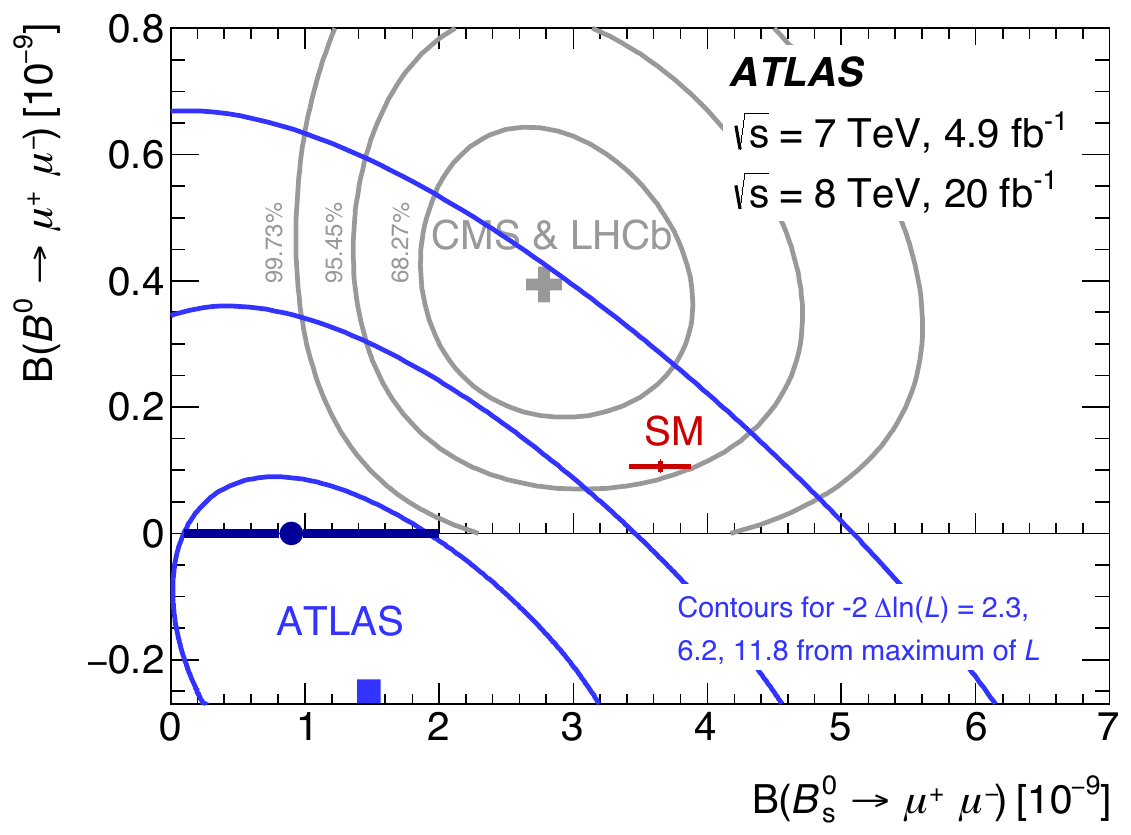}
\caption{ The two-dimensional likelihood scan result from ATLAS experiment. 
No significant signal excess found in both $B_s$ and $B^0\to\mu\mu$ channels. 
The results are still compatible with SM prediction (within $2\sigma$). }
\label{fig:bmm_atlas}
\end{figure}

As shown in Figure~\ref{fig:bmm_combine}, the $B_{s} \to \mu^+\mu^-$ channel has been first observed with a full combination of CMS and LHCb Run-1 data sets~\cite{CMS:2014xfa} with a significance of 6.2 standard deviations (7.4 expected). A mild hint of $B^0 \to \mu^+\mu^-$ decay has been seen. The combined branching fractions are given by
\begin{equation}
\mathcal{B}({B}_{s}\to\mu^+\mu^-) = (2.8^{+0.7}_{-0.6})\times10^{-9}~{\rm and}~\mathcal{B}({B^0}\to\mu^+\mu^-) = (3.9^{+1.6}_{-1.4})\times10^{-10}.
\end{equation}
The measured branching fraction of $B_s \to \mu^+\mu^-$ is in agreement with the SM prediction, while the $B^0 \to \mu^+\mu^-$ rate is slightly higher than the predicted value but still within the uncertainties. The ratio between $B^0$ and $B^0\to\mu^+\mu^-$ branching fractions has been measured as well; the major part of theoretical uncertainties cancelled and is a very clean test of the SM. 
ATLAS experiment has performed the corresponding measurement but no clear excess found~\cite{Aaboud:2016ire} (see Figure~\ref{fig:bmm_atlas}):
\begin{equation}
\mathcal{B}({B}_{s}\to\mu^+\mu^-) = (0.9^{+1.1}_{-0.8})\times10^{-9} (<3.0\times10^{-9})~{\rm and}~\mathcal{B}({B^0}\to\mu^+\mu^-) < 4.2\times10^{-10}.
\end{equation}
Although the expected significance is also modest and the results are compatible with the SM expectations. 

\begin{figure}[htb]
\centering
\includegraphics[height=2in]{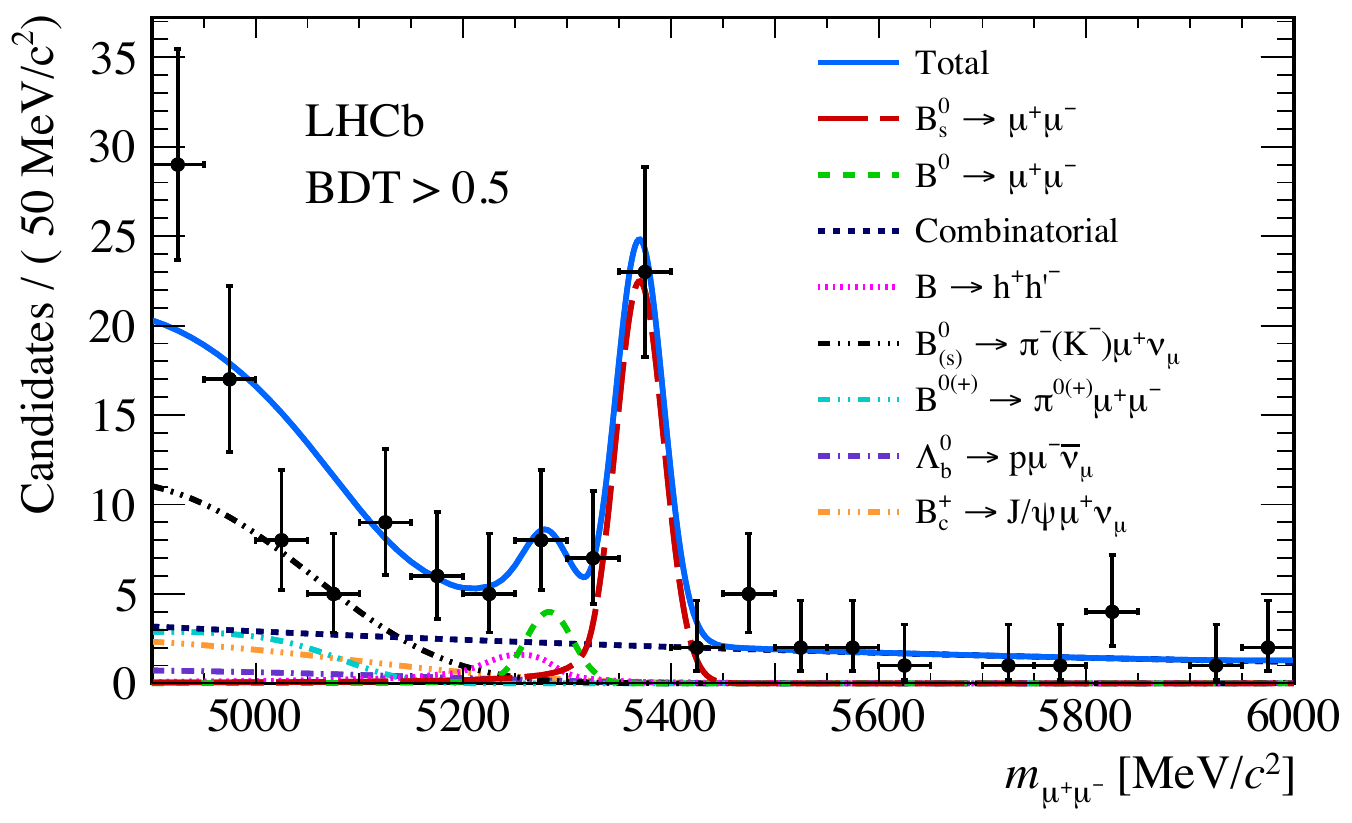}
\includegraphics[height=2in]{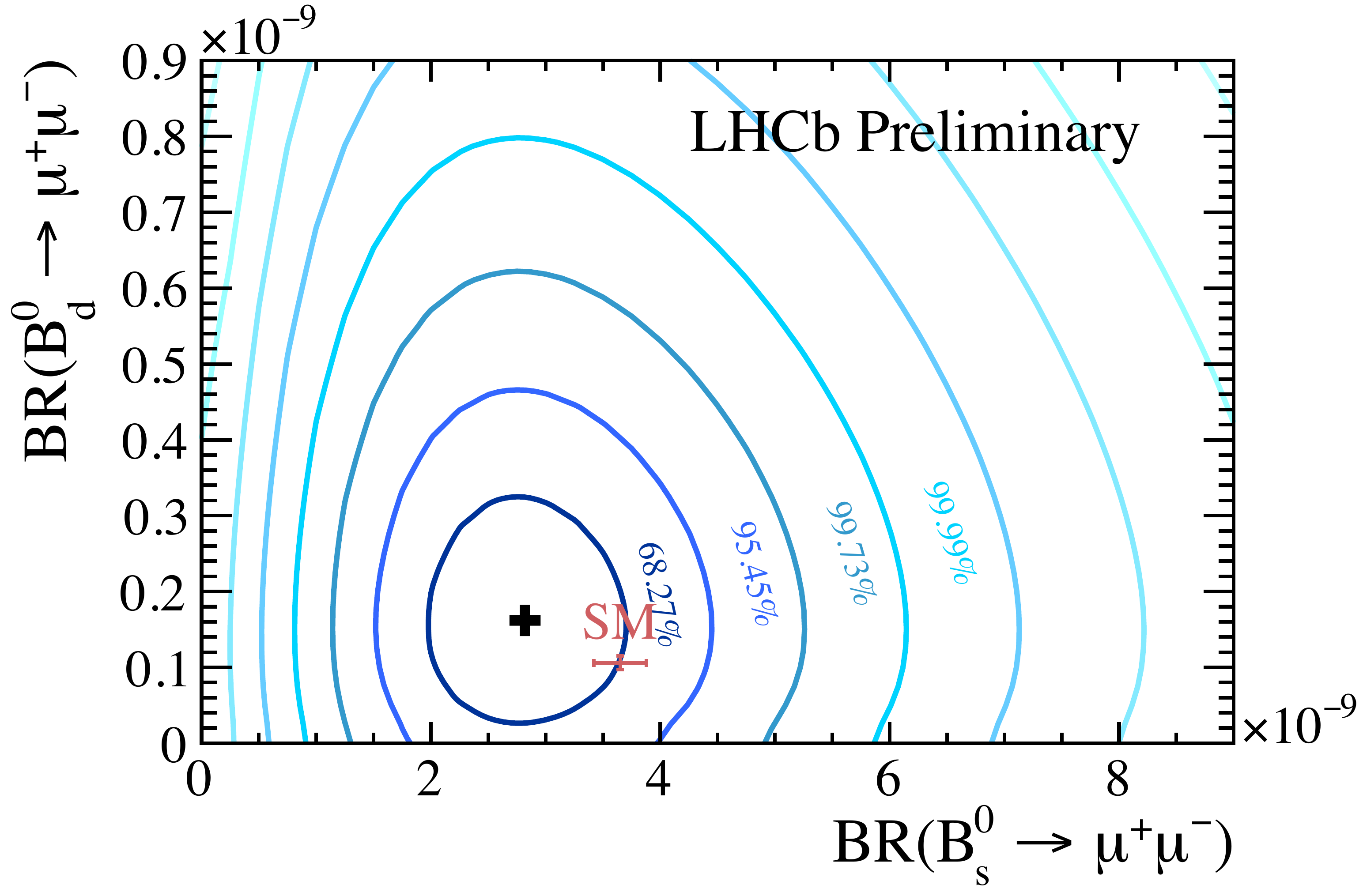}
\caption{ The invariant mass distribution for $B \to \mu\mu$ candidates (left) and the 
results of likelihood scans (right) from the full analysis of LHCb data. 
The results are consistent with the SM predictions as well as previous measurements.}
\label{fig:bmm_lhcb}
\end{figure}

Recently LHCb collaboration just updated the analysis by including 2011, 2012 and 2016 data sets~\cite{Aaij:2017vad}. 
With respect to the Run-I analysis, an improved isolation variable (BDT-based) has been introduced,
and new particle identification is adopted. Also a better BDT-based event classifier has been included.
As shown in Figure~\ref{fig:bmm_lhcb}, the analysis confirms 
the observation of $B_{s} \to \mu^+\mu^-$ decay at 7.8$\sigma$:
\begin{equation}
\mathcal{B}({B}_{s}\to\mu^+\mu^-) = (3.0\pm0.6^{+0.3}_{-0.2})\times10^{-9}~{\rm and}~\mathcal{B}({B^0}\to\mu^+\mu^-) < 3.4\times10^{-10}.
\end{equation}
The excess of $B^0 \to \mu^+\mu^-$ is reduced slightly and is in better agreement with the SM prediction.
With the same data set, LHCb also performed the first effective lifetime measurement with  ${B}_{s}\to\mu^+\mu^-$ events,
\begin{equation}
\tau({B}_{s}\to\mu^+\mu^-) = 2.05 \pm 0.44 \pm 0.05~{\rm ps},
\end{equation}
where ${B}_{s}\to\mu^+\mu^-$ only contributed by the heavier $B_s$ state in the SM.

Besides $B_{s},B^0 \to \mu^+\mu^-$ decays, LHCb also provided the first assessment for $B \to \tau^+\tau^-$ decays~\cite{Aaij:2017xqt}. 
In this channel the signal peak in mass distribution cannot be properly reconstructed, and it is not possible to distinguish $B_s$ and $B^0$ mesons.
So far there is no signal hint found in data. If one attributes all of the events to be either $B_s$ or $B^0$, the limits on the branching fractions 
can be derived. 
\begin{equation}
\mathcal{B}({B}_{s}\to\tau^+\tau^-) < 6.8\times10^{-3}~{\rm and}~\mathcal{B}({B^0}\to\tau^+\tau^-) < 2.1\times10^{-3}.
\end{equation}
A very new search for $K_S \to \mu^+\mu^-$ has been performed by LHCb as well~\cite{Aaij:2017tia}. No enhanced of signal found in data and an upper limit of
\begin{equation}
\mathcal{B}({K}_{S}\to\mu^+\mu^-) < 1.0 \times10^{-9}
\end{equation}
at the 95\% confidence level is presented.

\section{Angular analysis of $B^0 \to K^{*0} \mu^+\mu^-$}

The decays involving FCNC processes, such as $B^0 \to K^{*0} \mu^+\mu^-$ , are very sensitive to the models beyond the SM. New physics 
phenomena will contribute to these decays through the loop processes. For  $B^0 \to K^{*0} \mu^+\mu^-$ decay, one can define a set of observables
that can be measured in bins of $q^2$ (where $q^2$ is the dimuon invariant mass squared). The typical observables include 
the differential branching fraction ($d\mathcal{B}/dq^2$), the forward-backward asymmetry ($A_{FB}$), and longitudinal polarization fraction ($F_L$).
The predictions for these observables are robust, hence any difference between the experimental measurements and the predictions can be a smoking gun hint of new physics.

The target decay has four particles in the final state, $B^0 \to K^{*0} \mu^+\mu^- \to K^+\pi^-\mu^+\mu^-$. A full description of the decay requires three angular 
observables: $\theta_K$ (the helicity angle of the $K^{*0}$ candidate), $\theta_\ell$ (the helicity angle for the dimuon), and $\phi$ (the angle between the $K^{*0}$ and dimuon planes):
\begin{eqnarray} 
{d\Gamma\over dq^2 d\Omega} &= {9\over 32\pi}[
{3\over 4}(1-F_L) \sin^2\theta_K + F_L \cos^2\theta_K + {1\over 4}(1-F_L)\sin^2\theta_K\cos\theta_\ell \\
\nonumber & -F_L\cos^2\theta_K\cos2\theta_\ell + S_3 \sin^2\theta_K \sin^2\theta_\ell\cos2\phi \\
\nonumber & +S_4 \sin2\theta_K \sin2\phi_\ell\cos\phi + S_5 \sin2\theta_K \sin\theta_\ell\cos\phi \\
\nonumber & +{4\over 3} A_{FB} \sin^2\theta_K \cos\theta_\ell + S_7 \sin2\theta_K \sin\theta_\ell\sin\phi \\
\nonumber & +S_8 \sin2\theta_K \sin2\theta_\ell\sin\phi + S_9\sin^2\theta_K\sin^2\theta_\ell\sin2\phi],
\end{eqnarray}
where the observables $S_n$ are the bilinear combinations of the $K^{*0}$ amplitudes.
In this formula, the rates are averaged for different $CP$ states.

The B-factory experiments have reported some mild deviations in the measurement of $A_{FB}$ in the low $q^2$ region. The  
$A_{FB}$ is predicted to be mostly positive for higher $q^2$ region, while it goes to negative when $q^2$ approaches zero, and hence 
there is a zero-crossing point around 4~GeV$^2$. The early measurements from Belle and BaBar hint a positive $A_{FB}$ through out 
the whole $q^2$ phase-space, and it is challenging  to explain this anomaly within the framework of SM. However this anomaly
is not confirmed by the high statistic LHC measurements. All of the recent $A_{FB}$ measurements are consistent with the 
SM predictions.

Several new observables are proposed to have a better probe of the $B^0 \to K^{*0} \mu^+\mu^-$ decays. These new observables, 
which are denoted as $P^\prime_4$, $P^\prime_5$, $P^\prime_6$ and $P^\prime_8$~\cite{Descotes-Genon:2013vna}, are designed to 
have little dependency on the form factor uncertainties, in particular in the low $q^2$ range.
The $P^\prime_n$ observables are the combinations of $F_L$ and $S_n$:
\begin{equation}
P^\prime_{4,5,8} = {S_{4,5,8} \over \sqrt{F_L(1-F_L)}}~{\rm and}~P^\prime_{6} = {S_{7} \over \sqrt{F_L(1-F_L)}}.
\end{equation}
The measurements of these observables
are first performed by the LHCb collaboration and some renewed discrepancy between the measurements and the SM-based predictions
in the $P^\prime_5$ observable is seen.

\begin{figure}[htb]
\centering
\includegraphics[height=1.4in]{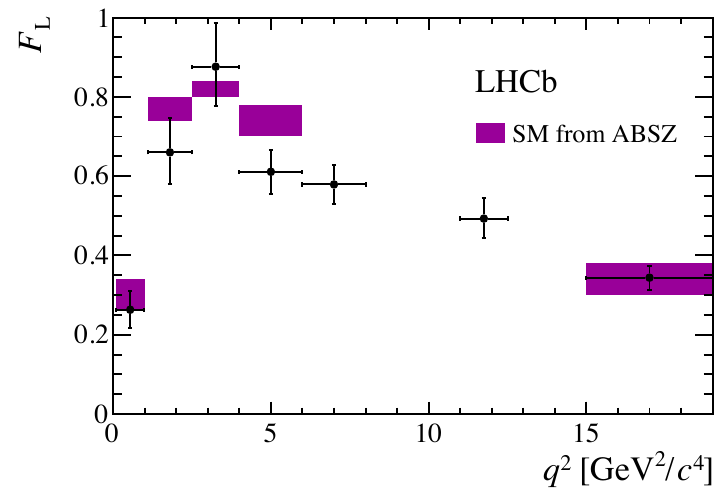}
\includegraphics[height=1.4in]{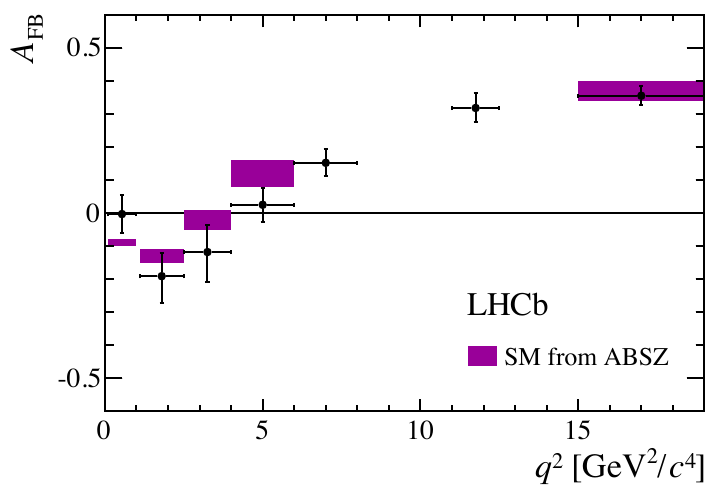}
\includegraphics[height=1.4in]{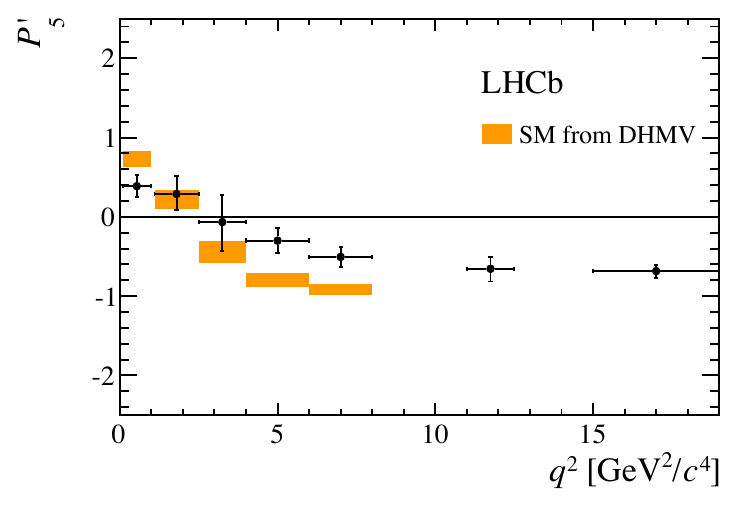}
\caption{ 
The longitudinal polarization fraction $F_L$ (left), 
the forward-backward asymmetry $A_{FB}$ (middle),
and the observable  $P^\prime_5$ (right) as measured in bins of $q^2$ from LHCb data. 
Many of the measured parameters (including $F_L$ and $A_{FB}$) are consistent with the SM predictions, while
a $3.4\sigma$ deviation found with respect to the SM from a global $\chi^2$ analysis.
}
\label{fig:kstmm_lhcb}
\end{figure}

A full angular analysis is required to measure the angular observables introduced above. 
The LHCb collaboration has performed the measurements~\cite{Aaij:2015oid} with two different statistical methods: a likelihood fit and a principle moments analysis.
As presented in Figure~\ref{fig:kstmm_lhcb},
the measured $P^\prime_5$ observable shows some deviation between the measured value and the SM-based prediction (DHMV~\cite{Descotes-Genon:2014uoa}) in the range of $4 < q^2 < 8$~GeV$^2$. From a global $\chi^2$ analysis, the statistical significance of the overall deviation 
is at 3.4$\sigma$. One can still explain this deviation by introducing some large hadronic effects within the SM, or new physics contributions
have to be added. 

ATLAS and CMS have recently produced the measurement of $P^\prime_n$ observables by reanalyzing the Run-I data sets~\cite{ATLAS:2017dlm,CMS:2017ivg}. 
Unlike LHCb and Belle, CMS and ATLAS have a difficulty to separate $K^{*0}$ and $\overline{K}^{*0}$, and the kaon flavour has to be chosen based on the best mass combination. This gives a small dilution to the measured observables. Also, in order to improve 
the fitting convergence, CMS and ATLAS analyses have to fold the signal model based on the symmetries to reduce the number of free parameters.
The CMS data is in better agreement with the SM-based ``DHMV" predictions. In the target $q^2$ region there is no 
significant deviations. The ATLAS result has a good agreement with SM predictions, except some deviations (of 2.5--2.7$\sigma$) from DHMV in 
the 4-6 GeV$^2$ $q^2$ region. Nevertheless, the statistical uncertainties are still too large and it is still early to draw a conclusion.
The most up-to-date measurements of $P^\prime_n$ observables from Belle experiment~\cite{Wehle:2016yoi} include both muon and electron channels. The muon result is consistent with the LHCb data, and shows a small discrepancy of 2.6$\sigma$ from the prediction. Interestingly, electron channel is in better agreement with the SM. 

\begin{figure}[htb]
\centering
\includegraphics[height=2in]{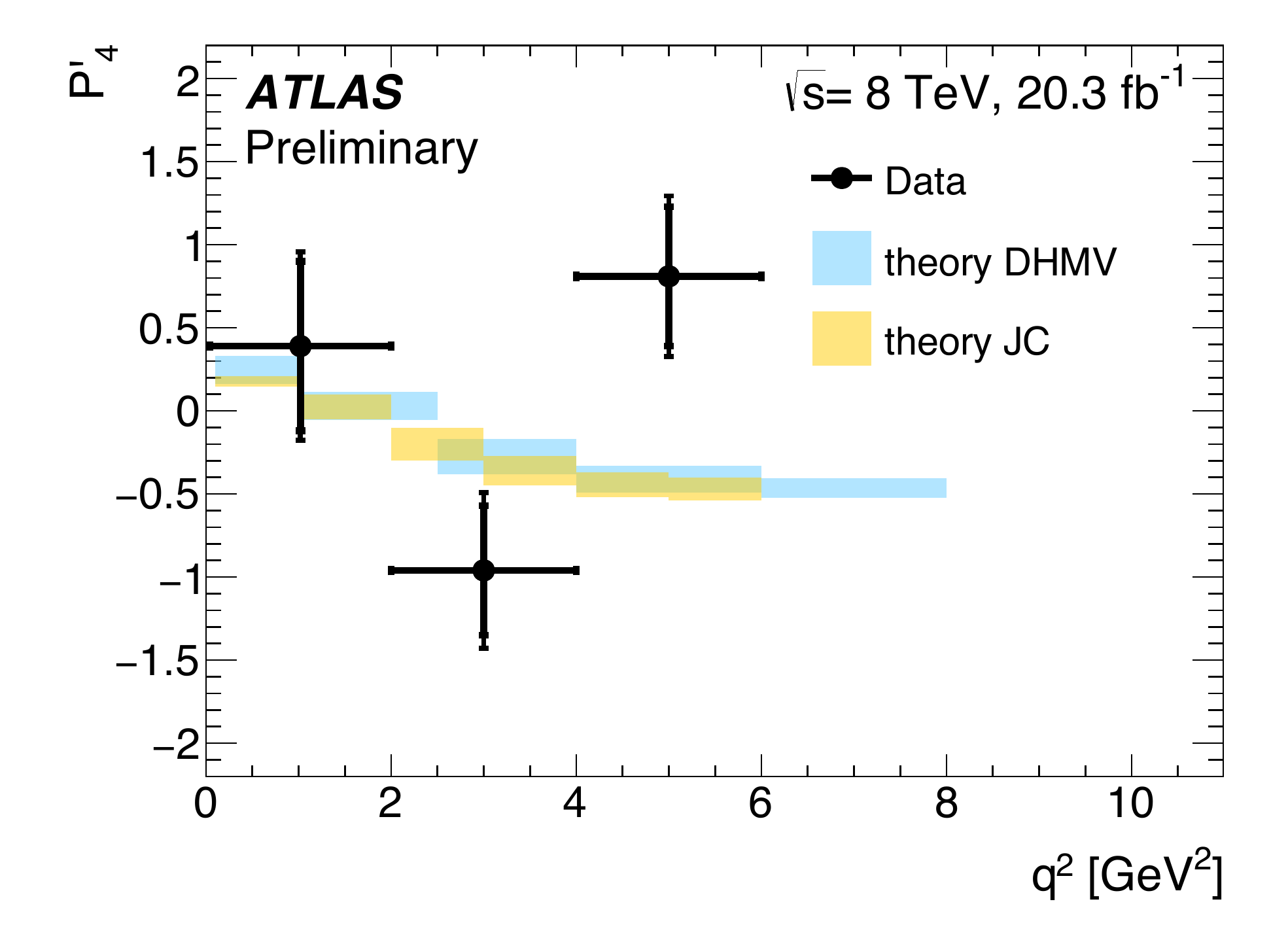}
\includegraphics[height=2in]{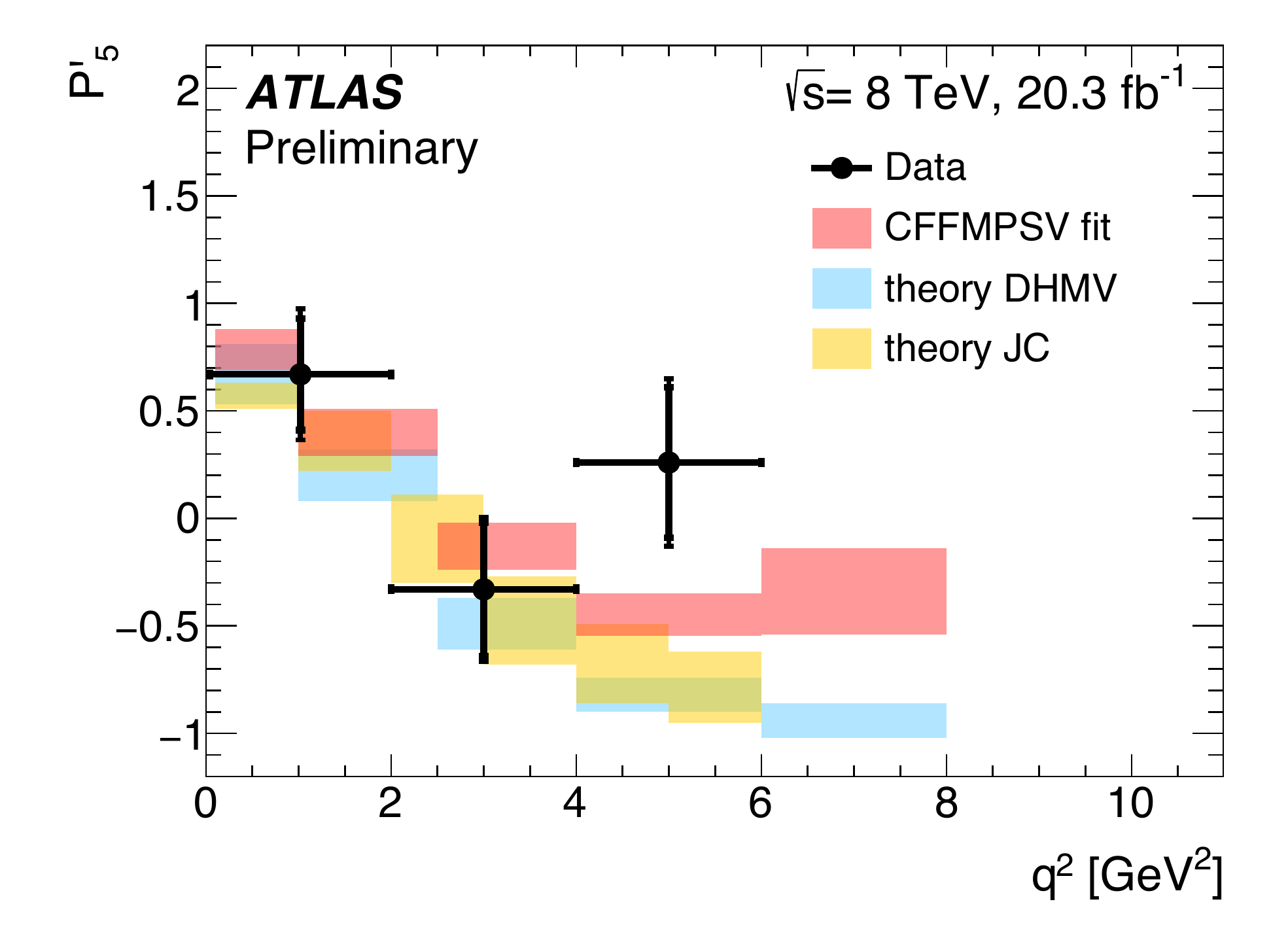}
\caption{ 
The observable  $P^\prime_4$  (left) and $P^\prime_5$ (right) as measured in bins of $q^2$ from ATLAS data. 
The results are in good agreement with the SM predictions, however some minor deviations of 2.5--2.7$\sigma$ from 
the DHMV expectations in the $4 < q^2 < 6$ GeV$^2$ region.
}
\label{fig:kstmm_altas}
\end{figure}

\begin{figure}[htb]
\centering
\includegraphics[height=2in]{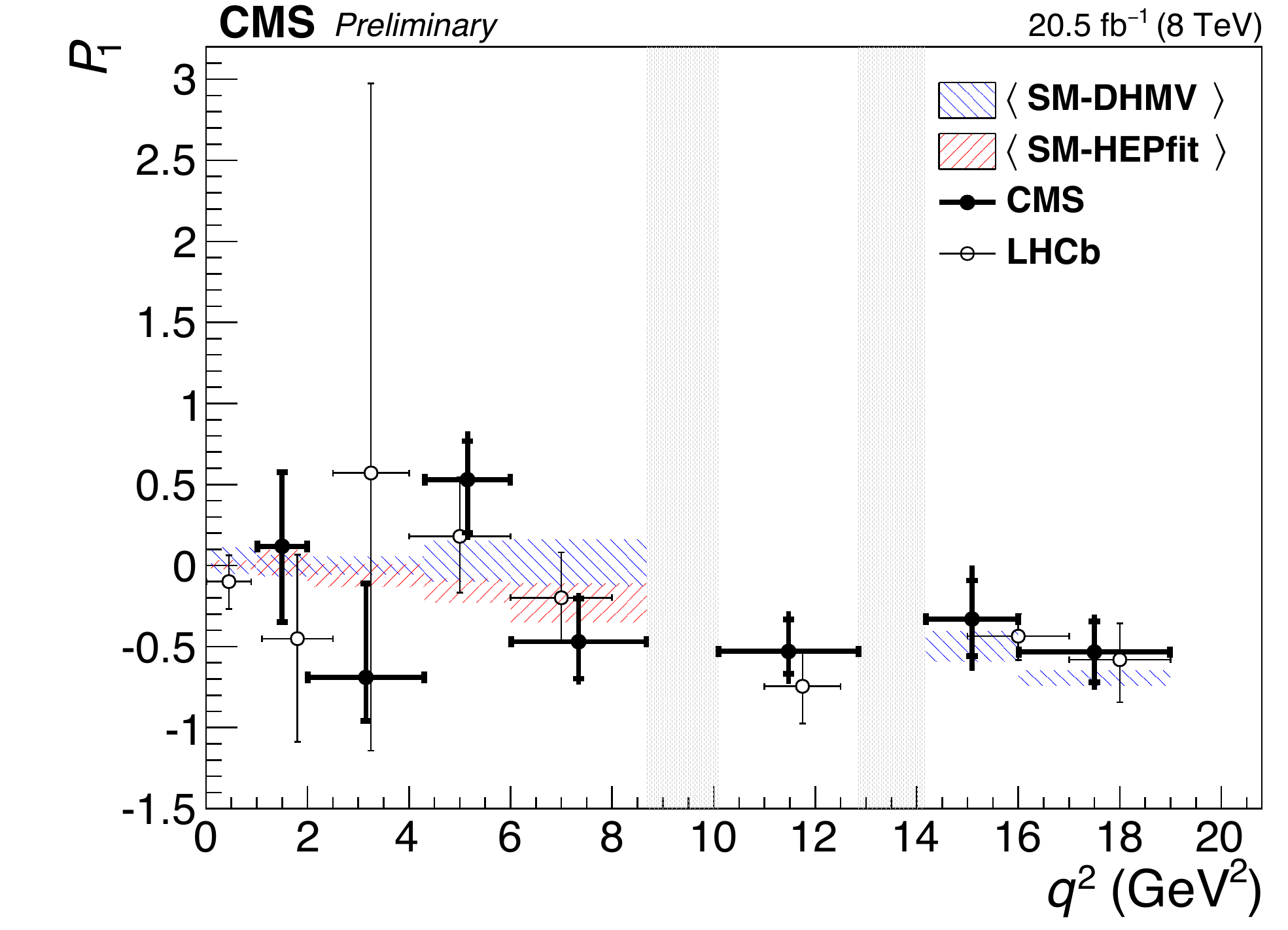}
\includegraphics[height=2in]{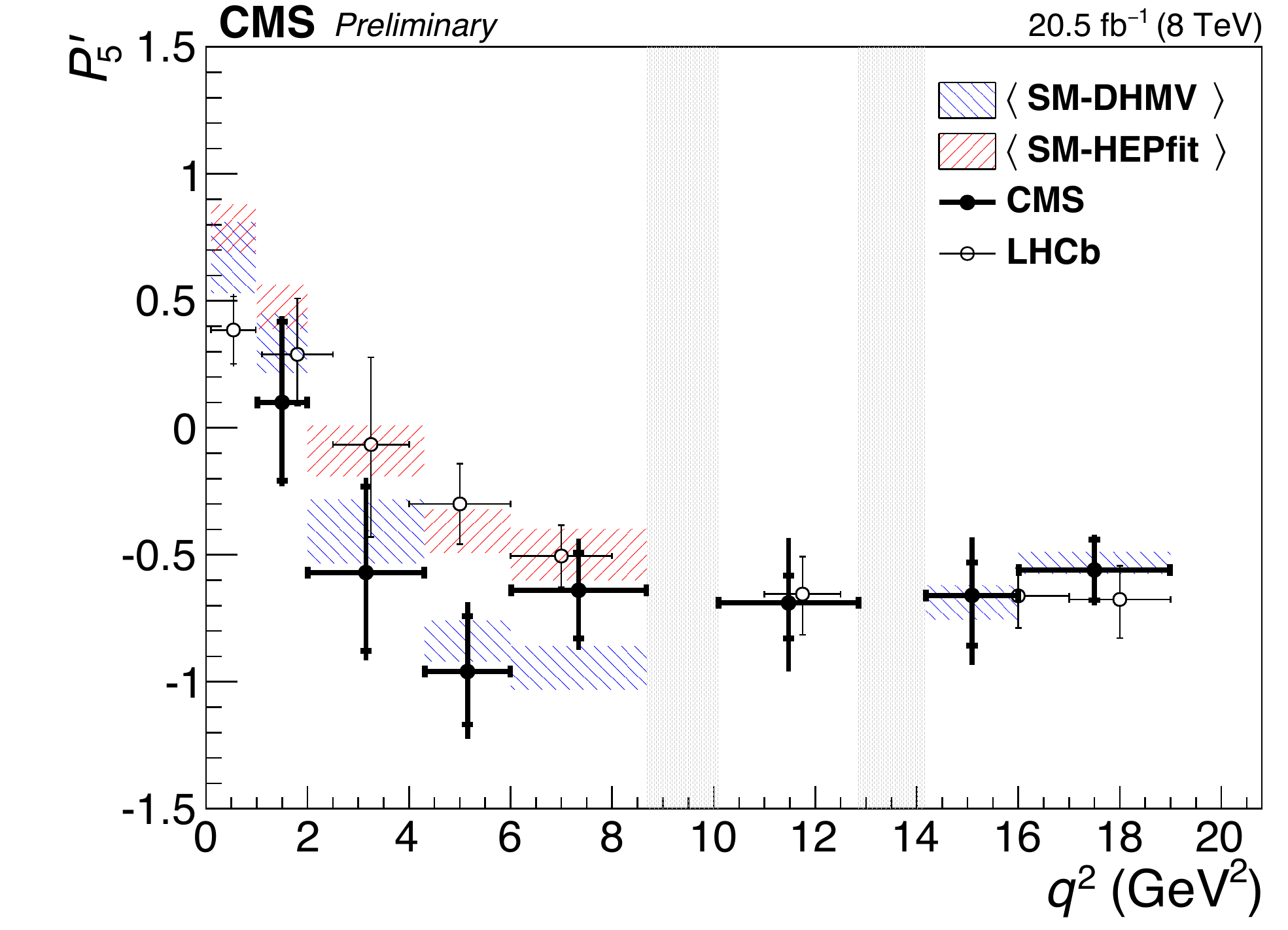}
\caption{ 
The observable  $P_1$  (left) and $P^\prime_5$ (right) as measured in bins of $q^2$ from CMS data. 
No significant deviation from the DHMV expectations.
}
\label{fig:kstmm_cms}
\end{figure}

As a summary remark, the deviations in $P^\prime_n$ observables are not the only place that data shows some unexpected tendency. 
In many of related $b\to s\ell\ell$ processes, the differential branching fractions measured by LHCb and CMS (see Ref.~\cite{Aaij:2014pli,Aaij:2015xza,Aaij:2016flj,Khachatryan:2015isa}) are generally found to be lower than the predictions in low $q^2$ region.
Some further investigation is required to identify this as a contribution from the hadronic effect, or something really new.
It is expected to have improved measurements from upcoming LHC data and from Belle-II in near future.

\end{document}

%% file: econfmacros.tex
\def\beq{\begin{equation}}
\def\eeq#1{\label{#1}\end{equation}}
\def\eeqn{\end{equation}}

\def\beqa{\begin{eqnarray}}
\def\eeqa#1{\label{#1}\end{eqnarray}}
\def\eeqan{\end{eqnarray}}

\let\bar=\overbar

\def\Dslash{\not{\hbox{\kern-4pt $D$}}}
\def\dslash{\not{\hbox{\kern-2pt $\del$}}}

\def\msb{{\bar{\ssstyle M \kern -1pt S}}}

